\documentclass[9pt]{book}

\usepackage[dvips]{graphicx,color}
\usepackage{makeidx,far,detector}

\usepackage{amsthm}
\usepackage{amstext}
\usepackage{amsmath}
\usepackage{amsbsy}
\usepackage{amssymb}
\usepackage{amsfonts}

\BookTitle{Far Detector in Korea for the J-PARC Neutrino Beam}
\CopyRight{\copyright 2007 by Universal Academy Press, Inc.}

\begin{document}
\pagenumbering{arabic}

\chapter{%
Neutrino Oscillations With A Next Generation Liquid Argon TPC
Detector in Kamioka or Korea Along The J-PARC Neutrino Beam}

\author{\raggedright \baselineskip=10pt%
{\bf A.\ Meregaglia,$^{1}$
and
A.\ Rubbia$^{1}$}\\ 
{\small \it %
(1) Institut f\"{u}r Teilchenphysik, ETHZ, CH-8093 Z\"{u}rich,
Switzerland
}
}






     \baselineskip=10pt
     \parindent=10pt

\section*{Abstract}
The ``baseline setup'' for a possible, beyond T2K, 
next generation long baseline experiment along the J-PARC neutrino
beam produced at Tokai, assumes two very large deep-underground Water Cerenkov imaging detectors
of about 300~kton fiducial each, located one
in Korea and the other in Kamioka but at the same off-axis angle. 
In this paper, we consider the physics performance of 
a similar setup but with a single and smaller, far detector, possibly at shallow depth, composed of 
a 100~kton next  generation liquid Argon Time Projection Chamber.
The potential location of the detector could be in the Kamioka area ($L\sim 295$~km)
or on the Eastern Korean coast ($L\sim 1025$~km),
depending on the results of the T2K experiment. In Korea the off-axis angle
could be either $2.5^{o} \sim 3^{o}$ as in SuperKamiokande, or
 $\sim 1^{o}$ as to offer pseudo-wide-band beam conditions.

\section{Introduction}

The well-established observations of
solar and atmospheric neutrinos, in particular from Superkamiokande (SK)~\cite{Kajita:2006gs}, SNO~\cite{Ahmad:2002jz} 
and KamLAND~\cite{Eguchi:2002dm} and the recent negative
result from MiniBOONE\cite{AguilarArevalo:2007it}, are very strong indicators of
the validity of the $3\times 3$ PMNS~\cite{pontecorvo} mixing
matrix U ($\nu_\alpha = U_{\alpha i}\nu_i$)
to describe all the observed neutrino flavor conversion phenomena.
In order to complete this picture, all the elements (magnitude
and phase) of the mixing matrix must be determined. That includes the
$U_{e3}$ element ($|U_{e3}|^2=\sin^2\theta_{13}$
in the standard parameterization), 
for which today there is only an upper bound. The
best constraint comes
from the CHOOZ~\cite{Apollonio:1999ae} reactor experiment,
corresponding
 to  $\sin^22\theta_{13}\lesssim 0.14$ (90\%C.L.) at
 the mass squared difference
$|\Delta m^2_{atm}|\simeq 2.5\times 10^{-3}\ \mathrm{eV}^2$.  

Long baseline neutrino experiments have the task to complete the present knowledge on 
the mixing parameters, possibly including the complex phase.
A non vanishing $|U_{e3}|$ would open the possibility of CP/T violation
in the leptonic sector, as a direct consequence of non-trivial complex phases 
in the $3\times 3$ mixing matrix. 
The determination of the missing elements is possible via the study of
$\nu_\mu\rightarrow\nu_e$ transitions at a baseline $L$ and energy $E$ 
with the choice $E/L \sim \Delta m^2_{atm}$.  
At the same time, the sign of 
the parameter $\Delta m^2_{atm}$ is unknown and this affects, in some
region of the parameter space which depends on the value of
$\sin^22\theta_{13}$, our ability to unambiguously detect CP-violation,
as we will illustrate it for our setup later on.
On the other hand, the sign of $\Delta m^2_{atm}$  could be determined 
by (1) precision measurement of the $\nu_\mu\rightarrow\nu_e$ oscillation probability
as a function of energy at sufficiently long
baselines (i.e. $L\gtrsim 300$~km), (2) by a comparison
of oscillations involving neutrinos and antineutrinos at sufficiently long
baselines, (3) by a comparison of probabilities
at similar energies with two detectors, one located at a shorter and one at
a longer baseline, or (4) by some combination of all the above methods (1)-(3).
Hence, for some region of the $\sin^22\theta_{13}$ angle, the determination
of CP-violation and the neutrino mass hierarchy must be addressed
simultaneously in order to avoid the mass hierarchy degeneracy problem.


\begin{table}[tb]
  \begin{center}
    \begin{tabular}{|l|c|c|c|c|c|c|c|}
         \hline
       & \multicolumn{3}{c|}{J-PARC} &  \multicolumn{4}{c|}{CERN SpS}\\   
       &   design & upgrade  & ultimate            &     CNGS  & + & 1& 2\\
       &        \cite{Itow:2001ee}           &  \cite{NP08}         & 
       \cite{Itow:2001ee}Ê&& \cite{Meregaglia:2006du} &\cite{Baibussinov:2007ea} & \cite{Baibussinov:2007ea} \\
              \hline
              Proton energy $E_p$ & \multicolumn{2}{c|}{30 GeV} & 40 GeV &  \multicolumn{4}{c|}{400 GeV}\\
               \hline
               $ppp (\times 10^{13})$ & 33 & 67 & $>67$ & 4.8 & 14& 4.8 & 15\\
               \hline
               $T_c$ (s) & 3.64 & 2 &  $<2 $ & 6 & 6 & 6&6\\
               \hline
               Efficiency &  1.0 & 1.0 & 1.0 & 0.55 &0.83& 0.8& 0.8\\
               \hline
               Running (d/y) & 130  & 130 & 130 & 220& 220 & 240 & 280 \\
               \hline
               $N_{pot}$ $/$ yr ($\times 10^{19}$) &  $100$ & 380 & $\simeq 700$ & 7.6 & 33 & 12 & 43.3 \\
               \hline
               Beam power (MW) & 0.6 & 1.6 & 4  &    0.5 & 1.5 & 0.5 & 1.6\\
               \hline
               $E_p\times N_{pot}$  & 4 &11.5 & 28 & 3 & 13.2 & 4.7& 17.3\\
               ($\times 10^{22}$ GeV$\cdot $pot/yr) & & & & & & &  \\
              Relative increase & & $\times 3$ & $\times 7$ & & $\times 4$ & $\times 1.5$ & $\times 6$ \\
               \hline
               Timescale & $>2009$ & 2014? & $>$2014? & $>2008$ &  \multicolumn{3}{c|}{$>$2016 ?}  \\
               \hline
      \end{tabular}
  \caption{Design parameters for the various beams at J-PARC. Comparison with the 
  dedicated CNGS intensity and assumed upgrades of the CERN SpS 
  complex (see text). The beam power corresponds to the instantaneous power
  on the neutrino target while the product $E_p\times N_{pot}$ corresponds
  to the total amount of energy deposited on the target per year, which is more
  relevant to calculate neutrino event rates.}
  \label{tab:accelerators}
  \end{center}
  \end{table}



The purpose of this workshop was to investigate the possible options with an upgraded
J-PARC neutrino beam after the present T2K experiment~\cite{Itow:2001ee}, 
in particular by placing a second detector in Korea. 
The center of the T2K neutrino beam will go through underground beneath SK, and 
will automatically reach the sea level east of the Korean shore. 
Therefore,
placing a detector in an appropriate location in Korea will 
probe neutrino oscillations
at a baseline of 1000 to 1200 km away from the source.

 The neutrino beam spectrum in Korea will depend on the off-axis angle and
 on the exact geographical location chosen, because of the non-cylindrical
shape of the decay tunnel in the neutrino beam line. 
When the upper side of the beam at $2^\circ$ to $ 3^\circ$ off-axis angle is observed at SK,
the lower side of the same beam at $0.5^\circ$ to $3.0^\circ$ off-axis angle
can be observed in Korea.
See Ref.~\cite{ourbeamtalk}
for beam spectra computed in some reference locations and for details.

The measurements in T2K might
indicate that the $\theta_{13}$ angle is in a region where the 
simultaneous determination of neutrino 
mass hierarchy and the CP violating phase becomes possible.
In Ref.~\cite{Ishitsuka:2005qi},  the possibility of using two next generation $\sim 300$~kton fiducial
identical Water Cerenkov detectors placed at different baseline distances but
at the same off-axis angle
was explored. The authors concluded that this setup would have the ability
to resolve the mass hierarchy by comparing the event yields at the two baselines,
one being more sensitive to matter effects than the other. 

In the present paper, we consider an alternative setup composed of a single
100~kton liquid Argon TPC located somewhere on the eastern coast of Korea
with off-axis angles ranging from $\sim 1^{o}$ to 2.5$^{o}$. We
explore the physics reach of this experiment in terms of sensitivity to
discover $\theta_{13}$, to determine the CP-violation phase $\delta$ and
to resolve the mass hierarchy.

\section{The liquid Argon Time Projection Chamber}
The liquid Argon Time Projection Chamber (LAr TPC) (See Ref.~\cite{t600paper}
and references therein) is
a powerful detector for uniform and high accuracy imaging of massive active volumes. 
It is based on the fact that in highly pure Argon, ionization tracks can be drifted
over distances of the order of meters. 
Imaging is provided by position-segmented electrodes at the end of the drift path, continuously recording the 
signals induced. $T_0$ is provided by the prompt scintillation light. 

In this paper, we assume a detector following the 
GLACIER concept~\cite{Rubbia:2004tz} with mass order of 100~kton.
The pros and cons of the LAr TPC, in particular in comparison to the Water
Cerenkov Imaging technique,  can be summarized as follow:
\begin{itemize}
\item {\bf Pros}: The liquid Argon TPC imaging should offer optimal conditions
to reconstruct with very high efficiency the electron appearance signal in the energy
region of interest in the GeV range, while considerably 
suppressing
the NC background consisting of misidentified $\pi^0$'s. MC studies show
that an efficiency above 90\% for signal can be achieved while suppressing $NC$ background
to the permil level~\cite{t2kprop}. 
This MC result was shown to be
true over a wide range of neutrino energy, typ. between 0 and 5~GeV. 
If verified experimentally, this implies that the intrinsic $\nu_e$ 
flux will be the dominant
background in a liquid Argon TPC coupled
to a superbeam~\cite{Meregaglia:2006du}. The systematic error
on this flux ($\lesssim 5\%$) will be determining the final sensitivity of the experiment.

\item {\bf Pros}: the physics performance per unit mass of the LAr TPC
is expected to be superior to that of the WC detector; hence, the LAr TPC detector
could be smaller than the WC detector to achieve the same
physics performance; the 100~kton detector considered here is
approximately twice the size of the
Superkamiokande detector.
In addition, a LAr TPC should allow operation at shallow
depth. The
constraints on the excavation and the related siting issues of the detector
should hence be reduced in the case of a LAr TPC.
For a quantitative discussion on the possible
 operation at shallow depth,
see Ref.~\cite{Bueno:2007um}.

\item {\bf Pros}: the imaging properties and the good energy resolution
of the LAr TPC would allow to consider all events around the GeV region
and above, while the WC technology is essentially limited to quasi-elastic
(QE) events and background considerations limit the beam energy
to lie below the GeV range. Hence, 
broader band beams, as for example obtained at J-PARC with
smaller off-axis angles than SK, e.g. $1^{o}$~off-axis, covering more
features of the oscillation probability (e.g. first maxima, first minima,
second maxima, etc.) can be
contemplated.

\item {\bf Cons}: the community has less experience with the LAr TPC technology 
than the WC; the largest detector
ever operated, the ICARUS T300, has a modular design which is not easily
extrapolated to the relevant masses. Significant R\&D and improvements
in the design are therefore
required in order to reach a scalable design which could offer a path for
a 100~kton mass facility in a cost effective way.
For an overview and results of one such R\&D program, see 
Refs.~\cite{Ereditato:2005ru,Ereditato:2005yx,Badertscher:2005te,Rubbia:2005ge}. 

\item {\bf Cons}: the procurement and underground
handling of large amounts of liquid Argon is more difficult 
than that for water, however, safe, surface or near-surface,
storage of very large amounts of cryogen (with volumes larger
than the ones considered here) has been achieved by the petrochemical industry; liquid Argon
is a natural by-product of air liquefaction which has large industrial and commercial applications
and can be in principle produced nearby any chosen location.
\end{itemize}

\section{The J-PARC neutrino beam}
The measurements considered by future long baseline experiments will require to accumulate sufficiently large
statistics in the far detectors to study with high precision the neutrino oscillation
phenomena. 
The physics performance  of these facilities will therefore depend
on the ability of the involved accelerator complexes and of the neutrino beam infrastructures
to offer stable and long-term fault-less operations. In Table~\ref{tab:accelerators} we summarize
the design parameters for the J-PARC beam under the three conditions ``baseline'',
``upgraded'' and ``ultimate'':
\begin{itemize}
\item {\bf The ``baseline'' 0.6~MW T2K beam~\cite{Itow:2001ee}:}
The J-PARC neutrino beam is under
construction for the T2K experiment and is planned to begin
operation in 2009 at low intensity.  The final goal for the T2K experiment is to 
reach an integrated intensity of $5 \times 10^{21}$ pots,
or equivalently a beam power of $\sim 0.6$~MW during 5 years.

\item {\bf The ``upgraded'' 1.66~MW beam~\cite{NP08}:}
there is a plan to further upgrade the accelerator complex to 
potentially provide an increased
beam power of 1.66~MW to the neutrino target. This upgrade 
should in principle not require major modifications
in the beamline infrastructure which has been designed
up to 2~MW.

\item {\bf An ``ultimate'' 4~MW beam~\cite{Itow:2001ee}:}
The physics reach of  a megaton (Mton)-class water Cerenkov detectors
coupled to a  beam power of 4MW was originally considered
in Ref.~\cite{Itow:2001ee}. Although such a power might require significant
upgrades of the J-PARC facility, it was chosen as default value in the
``baseline setup'' for the two Water Cerenkov detectors configuration in Korea 
and Kamioka~\cite{Ishitsuka:2005qi}. For ease of comparison, we
also assume this value, unless otherwise stated.

\end{itemize}
Finally for reference, Table~\ref{tab:accelerators} includes the intensity expected at the
CERN NGS as well as figures corresponding to assumed upgrades
of the CERN accelerator complex~\cite{Meregaglia:2006du,Baibussinov:2007ea}. 
The upgraded CNGS performance will be 
discussed in Section~\ref{sec:cngs}. 

\section{Expected event rates in Korea and physics analysis}
We carry out an analysis following our previous work outlined in Ref.~\cite{Meregaglia:2006du}. 
For the fit procedure, we assume at this stage identical running periods for each horn polarity,
i.e. assume equal neutrino and antineutrinos runs. In an actual experiment,
the plain exploration of $\sin^22\theta_{13}$ would suggest more neutrinos, while
the CP-violation search requires a comparison between neutrinos and antineutrino
runs.
Assuming that the neutrino flux at the far location will be precisely known by extrapolation
of the near data measured by detectors located at 280~m~\cite{Itow:2001ee} and 
possibly at 2~km~\cite{t2kprop}, we assumed 
a systematic error 
on the $\nu_e+\bar\nu_e$ flux at the far location of $5\%$. 
The simulated data and background is fitted in energy bins of 100~MeV each.
More refined treatments of the systematic errors,
including also instrumental effects, will be included at a later stage.

Assuming an integrated intensity of $35\times 10^{21}$~pots (or  5 years of running at 4~MW)
for each horn polarity mode (labelled neutrino and antineutrino
runs for a total of 10~years), the number of events expected neglecting flavor oscillations
are reported in Table~\ref{tab:KoreaEventsNoOsc} for three locations:
(1) 295~km, OA 2.5$^o$ corresponding to the Kamioka region;
(2) 1025~km, OA 1.0$^o$ corresponding to most east region of Korea;
(3) 1025~km, OA 2.5$^o$. In the Table, the interactions of neutrinos
and antineutrinos are separated, however, for the analysis we sum them,
as we do not consider the possibility to discriminate between the two on
an event-by-event basis. The higher rates are clearly expected at
295~km, while the $1/L^2$ dependence reduces the flux in Korea
by an order of magnitude. A significant fraction of the loss compared to Kamioka
(OA 2.5$^o$) can indeed be recovered by locating the Korean detector
at OA~1.0$^o$. In this case, the rate in Korea is approximately half that at Kamioka.
In addition, given the different kinematics in 3-body kaon decays, the beam
background ratio $\nu_e/\nu_\mu$ is more favorable at smaller OA angles.

Based on these arguments, one would conclude that an OA~1.0$^o$ in Korea
is favored compared to the OA~2.5$^o$. However, it was pointed out by
the authors of Ref.~\cite{Ishitsuka:2005qi} that the same OA angle at Kamioka
and Korea would allow to cancel certain systematic errors associated to the
beam. At this stage we have not attempted to quantify this further.

\begin{table}[htb]
  \begin{center}
    \begin{tabular}{|c|c|c|c|c|c|c|}
           \hline
         &  \multicolumn{3}{c|}{neutrino run} & \multicolumn{3}{c|}{antineutrino run} \\
         \hline

      Location	 & $\nu_{\mu}$CC&	$\nu_e$CC  & ($\nu_{e} + \overline{\nu}_e$) / & $\nu_{\mu}$CC& $\nu_e$CC  & ($\nu_{e} + \overline{\nu}_e$) / \\     
      &  ($\overline{\nu}_{\mu}$CC) & ($\overline{\nu}_e$CC)&  ($\nu_{\mu} +  \overline{\nu}_{\mu}$)&  ($\overline{\nu}_{\mu}$CC) & ($\overline{\nu}_e$CC)&  ($\nu_{\mu} +  \overline{\nu}_{\mu}$)\\
      \hline
          \multicolumn{7}{|c|}	{\bf{J-PARC - 40 GeV/c protons - T2K optics - 4MW} } \\
          \hline
      \bf{295 km} & & & & & &\\
       2.5 deg &205000 &	3619	&1.9 $\%$ & 27562 & 1225 &2.7 $\%$ \\
             (0-5 GeV) & (5970)& (416) & &(60404) & (1136)&\\

          \hline
      \bf{1025 km} & & & & & &\\
      $\sim$ 1 deg &81650 &	716 	&0.9 $\%$ & 9737 & 176 &1.1 $\%$ \\
             (0-5 GeV)&(3249) &(60) & &(24415) &(212) &\\

     \hline
      \bf{1025 km} & & & & & &\\
      2.5 deg &16980 &	300 	&1.9 $\%$ & 2283 & 101&2.7 $\%$ \\
             (0-5 GeV)&(495)& (34)& &(5003) &  (94)&\\
\hline

      \end{tabular}
  \caption{Number of events calculated for $35 \times 10^{21}$ p.o.t. ($7 \times 10^{21}$ pot per year $\times$ 5 years) for each polarity mode and a detector of 100~kton. A cut of 5~GeV has been set on neutrino energy.}
   \label{tab:KoreaEventsNoOsc}
  \end{center}
\end{table}

\begin{table}[bth]

\centering
\begin{tabular}{|c|c|c|ccccc|}
 \hline
&& \multicolumn{6}{|c|}{neutrino run}\\ 
&& $\nu_{\mu}$CC& \multicolumn{5}{c|}{$\nu_e$CC}  \\
Location &{ sin$^2$ (2$\theta_{13}$) }& + & \multicolumn{5}{c|}{+} \\
&= 0.002&  $\overline{\nu}_{\mu}$CC & \multicolumn{5}{c|}{$\bar{\nu}_{e}$CC} \\
&& no osc. & $\delta$=0 & $90^o$ & $270^o$ & $180^o$ & beam \\ 
\hline
\multicolumn{8}{|c|}{\bf J-PARC - 40 GeV/c protons - T2K optics - 4MW}\\
\hline
    \bf{295 km}  & & &  &  &  &  &   \\
     2.5 deg&Matter (n.h.)  & 210970 & 274 & 39 & 393 & 158 & 4035 \\ 
 (0-5 GeV)& & &  &  &  &  & $\sqrt{B} = 64$  \\
\hline
   \bf{1025 km}  & & &  &  &  &  &   \\
        $\sim$ 1 deg &Matter (n.h.)  & 85900  & 226 & 138 & 389 & 300 & 776 \\ 
 (0-5 GeV) & & &  &  &  &  &  $\sqrt{B} = 28$ \\
 \hline
  \bf{1025 km}  & & &  &  &  &  &   \\
        2.5 deg &Matter (n.h.) & 17475 & 94 & 60 & 126 & 92 & 334 \\ 
 (0-5 GeV)& & &  &  &  &  &  $\sqrt{B} = 18$ \\
\hline

\end{tabular}
\caption{Number of oscillated events calculated for a detector of 100~kton
and $35 \times 10^{21}$ pots with horns on neutrino polarity. The parameters used for the oscillation are the following: $\Delta m_{32}^2$ = 2.5$\times 10^{-3}$ eV$^2$, $\Delta m_{12}^2 = 7\times10^{-5}$ eV$^2$, tg$^2$ ($\theta_{12}$)= 0.45, sin$^2$ ($\theta_{23}$)= 0.5, sin$^2$ (2$\theta_{13}$)= 0.002, $\rho=2.8~$g/cm$^3$.}
\label{tab:oscillation}
\end{table}


\section{Sensitivity on $\theta_{13}$}
We report  in Table~\ref{tab:oscillation} the
expected number of oscillated events including matter effects
for a normal neutrino hierarchy
($\rho=2.8$~g/cm$^3$) for different values of the
CP-violation phase 
$\delta$ and $\sin^2 (2\theta_{13}) = 0.002$
for the three same detector locations of Table~\ref{tab:KoreaEventsNoOsc}.
The parameters used for the oscillation are the following: $\Delta m_{32}^2$ = 2.5$\times 10^{-3}$ eV$^2$, $\Delta m_{12}^2 = 7\times10^{-5}$ eV$^2$, tg$^2$ ($\theta_{12}$)= 0.45, sin$^2$ ($\theta_{23}$)= 0.5, 
and $\sin^2 (2\theta_{13})= 0.002$.

The number of oscillated events depends as expected on the chosen value of the $\delta$-phase.
For the normal hierarchy and neutrinos, the smallest number of events is observed for
$\delta$ around 90$^o$. Relative to the intrinsic beam $\nu_e$ background, the number of oscillated
events is significant (see table for the $\sqrt{B}$), 
even for the small value of  $\sin^2 (2\theta_{13}) = 0.002$.

\begin{figure} [tb]
\begin{center}
     \includegraphics[width=.8\textwidth]{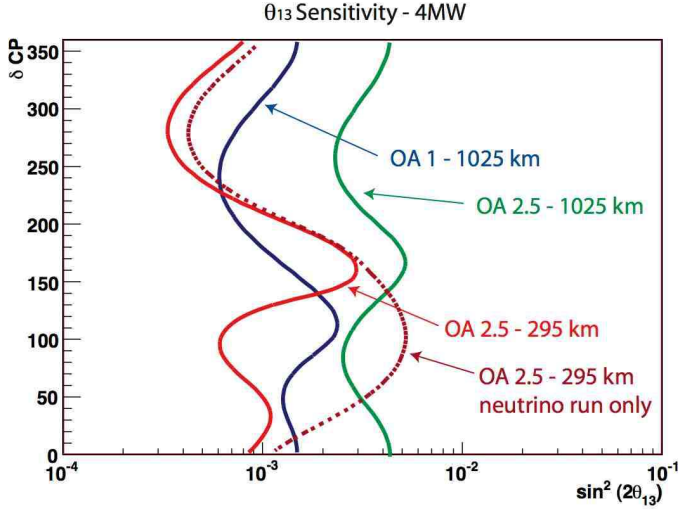}
\caption{$\theta_{13}$ sensitivity at $3\sigma$~C.L. for our three locations. 
The dashed line corresponds to a neutrino run only for the 295~km, OA 2.5$^o$  configuration.}
\label{fig:th13}
\end{center}
\end{figure}

These results translate in the $\theta_{13}$ $3\sigma$ C.L. sensitivity shown in Figure~\ref{fig:th13}. 
The configurations 295~km, OA 2.5$^o$ and  
 1025~km, OA 1.0$^o$ yield similar sensitivities, while 1025~km, OA 2.5$^o$ is
 slightly worse.
The role of the antineutrino run for the region $0 < \delta < 150^o$
 is also illustrated with the dashed curve which corresponds to
 a neutrino run only.

\section{CP-violation discovery}

For our definition of CP-violation sensitivity and the sensitivity fitting procedure
please refer to Ref.~\cite{Meregaglia:2006du}. 
In short, the CP-violation can be said to be
discovered if the CP-conserving values, $\delta=0$ and $\delta=180^o$, 
can be excluded at a given C.L. The reach for discovering CP-violation is computed
choosing a ``true'' value for $\delta$ ($\ne 0)$ as input at different true values
of $\sin^22\theta_{13}$ in the $(\sin^22\theta_{13},\delta)$-plane,
and for each point of the plane calculating the corresponding 
event rates expected in the experiment. This data is then fitted with
the two CP-conserving values $\delta=0$ and $\delta=180^o$, leaving
all other parameters free (including $\sin^22\theta_{13}$~!).
The opposite mass hierarchy is also fitted and the minimum of all
cases is taken as final $\chi^2$. 

\begin{figure} [tbh]
\begin{center}
     \includegraphics[width=.495\textwidth]{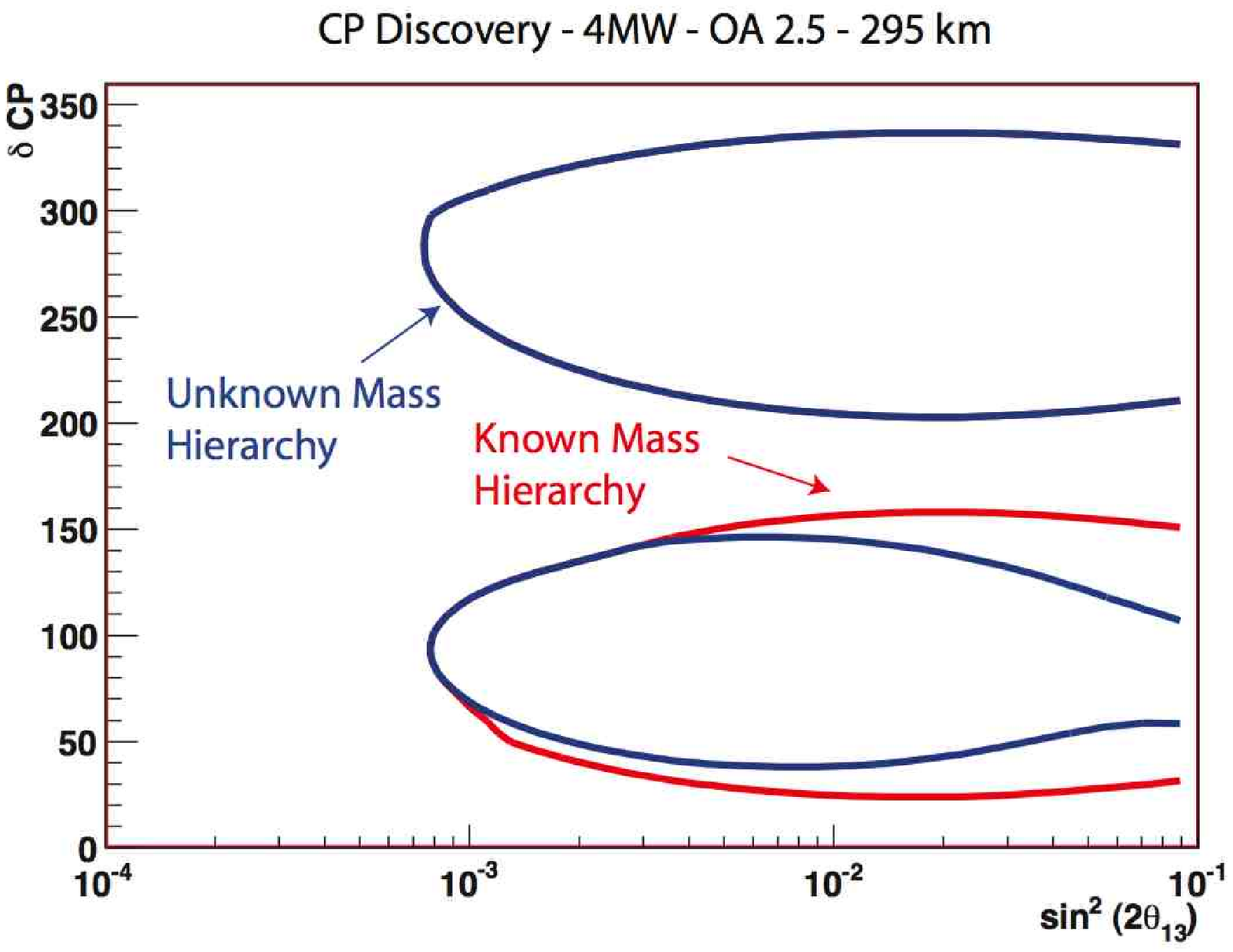}
     \includegraphics[width=.495\textwidth]{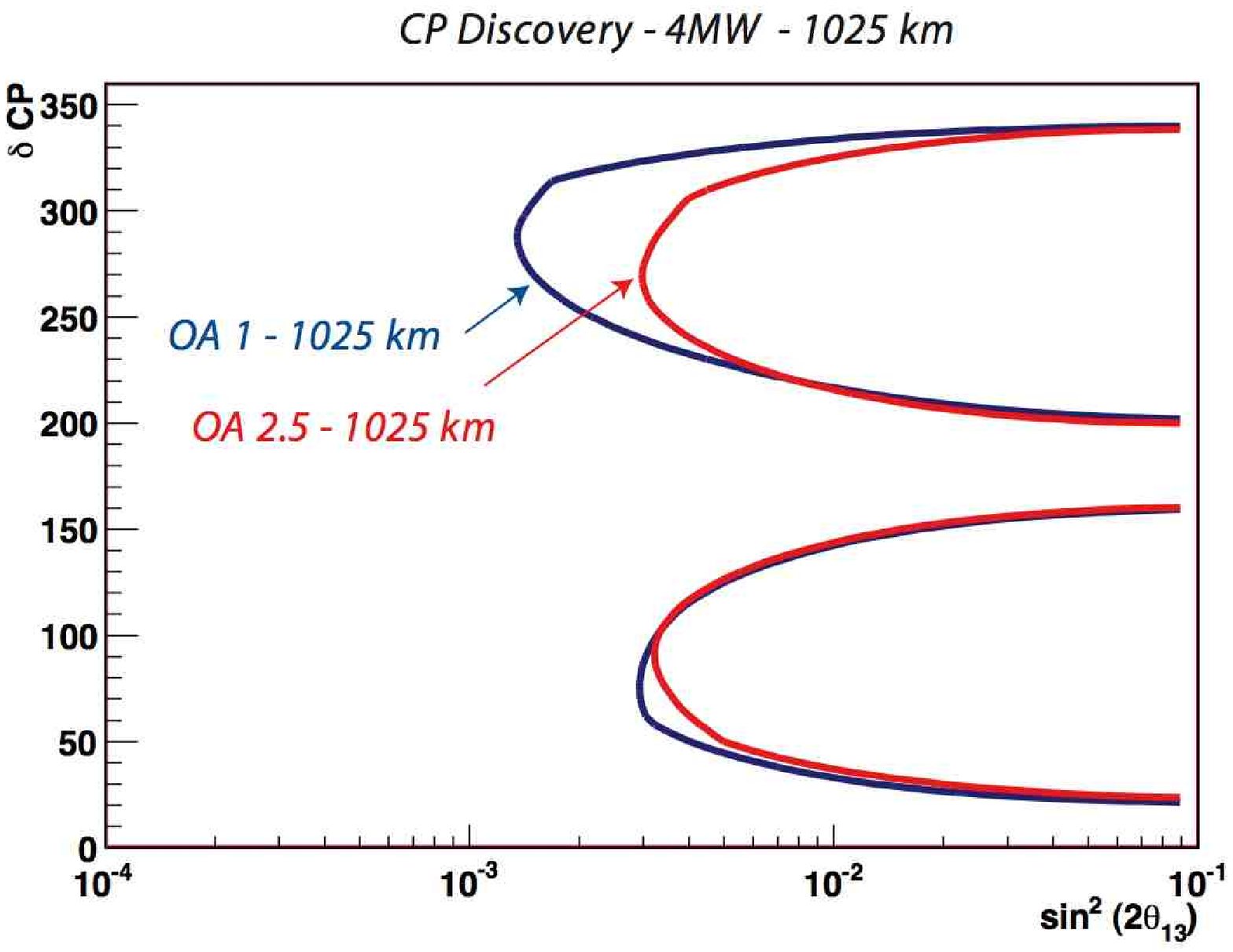}
\caption{CP-violation discovery at $3\sigma$~C.L. 
assuming an integrated intensity of $35\times 10^{21}$~pots
for each horn polarity mode (5 yrs neutrino and 5 yrs antineutrino
runs for a total of 10~years), for the (left) 295~km with (blue line) and without (red line)
a priori knowledge of the mass hierarchy and (right) 1025~km OA~1$^o$ (blue line)
and OA~2.5$^o$ (red line).}
\label{fig:CP}
\end{center}
\end{figure}

Leaving all unknown parameters free and letting vary the known ones within their experimental
errors in the fit, the CP-violation $3\sigma$ C.L. discovery sensitivity is shown in Figure~\ref{fig:CP}
for the three geographical configurations 295~km 
OA~2.5$^o$,  1025~km OA~1$^o$ and OA~2.5$^o$,
assuming an integrated intensity of $35\times 10^{21}$~pots
for each horn polarity mode (5 yrs neutrino and 5 yrs antineutrino
runs for a total of 10~years).
For the shortest baseline (295~km), the mass hierarchy degeneracy affects the sensitivity
at large $\sin^22\theta_{13}\gtrsim 10^{-2}$. Given the limited sensitivity
to matter effects at this baseline, the data could be fitted with a conserving value of $\delta$ and 
the wrong (opposite to true) sign of $\Delta m^2_{32}$.

This effect can be qualitatively understood by noting that the neutrino beam is optimised to cover the first maximum of oscillation which corresponds, at the selected baseline, to about 600 MeV neutrinos. At this energy the oscillation bi-probabilities for neutrinos and antineutrinos 
with $\delta = 90^o$ and $\Delta m^2>0$,
essentially coincide with that with
$\delta= 0$ and $\delta = 180^o$ for  $\Delta m^2<0$,
therefore the data can be very well fitted with a conserving value of $\delta$ and the opposite mass 
hierarchy.

The mass hierarchy degeneracy problem can be resolved by choosing a longer baseline and/or by 
extending the energy range of the beam spectrum, like for example is the case when we reduce
the OA angle.
In the Korean configuration, the situation is resolved as can be seen from Figure~\ref{fig:CP} (right):
the CP-violation discovery region for the OA 1 and OA 2.5 configurations are therefore
favored for large values of $\sin^22\theta_{13}$, but it also emerges
 that the small off-axis angle configuration is favoured by the higher statistics, 
 and for small values of $\theta_{13}$ the sensitivity is better than that obtained in Korea.

\begin{figure} [tbh]
\begin{center}
     \includegraphics[width=.8\textwidth]{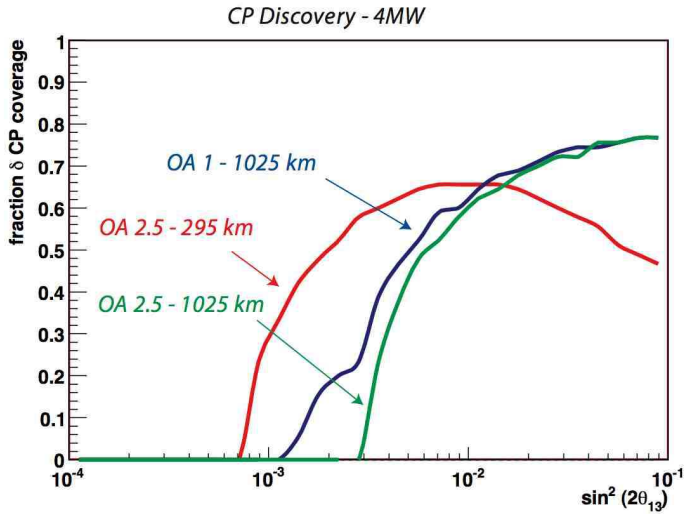}
\caption{Fraction of $\delta$ coverage for CP-violation sensitivity at $3\sigma$~C.L. for our
three configurations, corresponding to Figure~\ref{fig:CP}.}
\label{fig:CPcov}
\end{center}
\end{figure}

To summarize, the results are plotted 
in terms of $\delta$ coverage in Figure~\ref{fig:CPcov}. 
If the Korean scenario is chosen, the OA 1$^o$ angle guarantees a better sensitivity.
In the Kamioka scenario, the CP-coverage decreases for values of $\sin^22\theta_{13}\gtrsim0.01$.
It is a striking coincidence that this value approximately corresponds to the expected
sensitivity of the T2K experiment! Hence, 
in case a signal is observed with $\sin^22\theta_{13}\gtrsim0.01$ one should aim at the largest possible $\delta$ coverage, and a longer baseline should be chosen to avoid neutrino mass hierarchy degeneracy.
In this case, the authors of Ref.~\cite{Huber:2007em} claim that one should pay attention to other systematic errors.

On the contrary, if no signal is measured and $\sin^2 (2 \theta_{13})\lesssim 0.01$, the priority is to reach the best CP-sensitivity for the smallest possible value of the mixing parameter $\theta_{13}$ and the
shortest baseline could be favored.

\section{Mass hierarchy determination}

\begin{figure} [tbp]
\begin{center}
     \includegraphics[width=.8\textwidth]{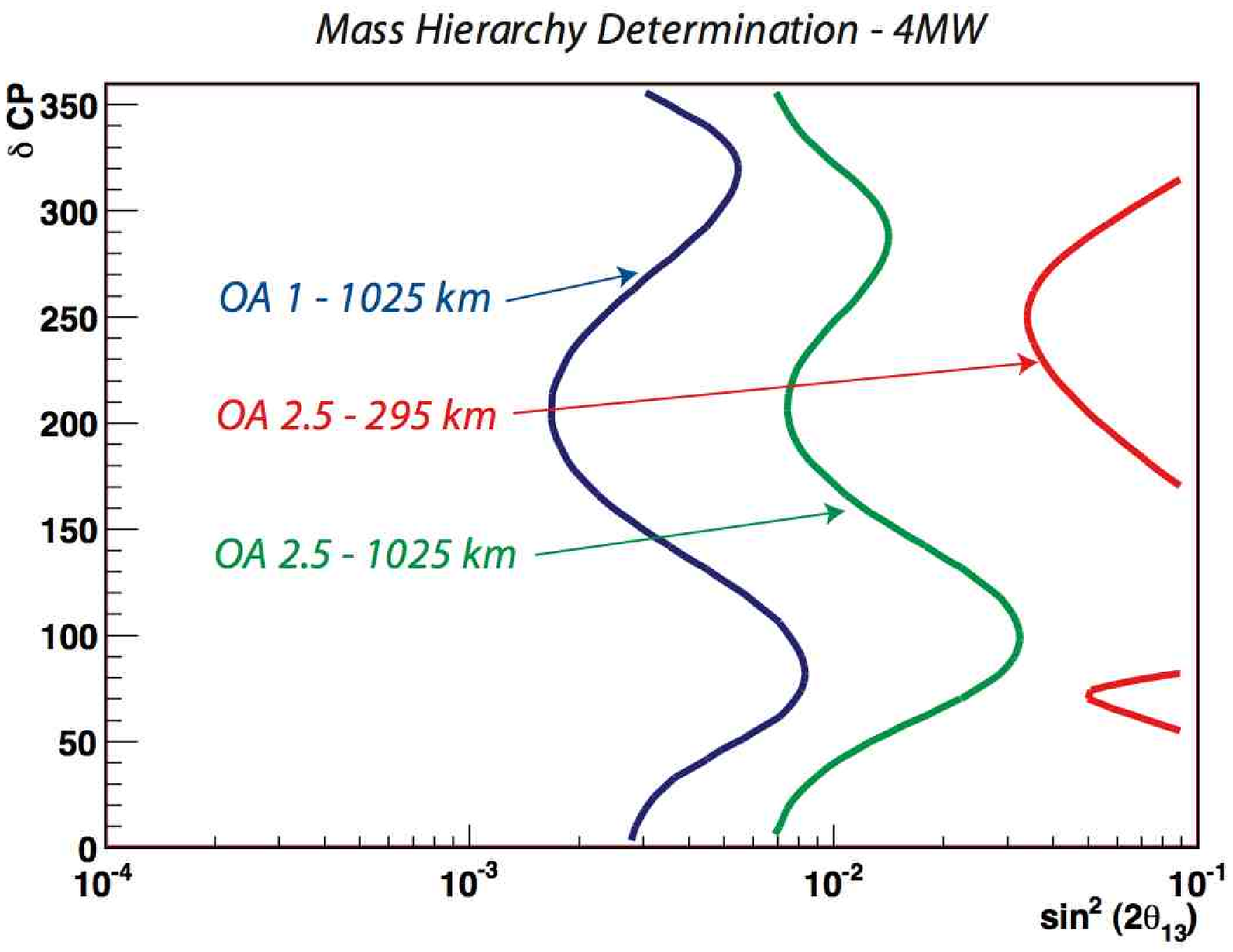}
\caption{Mass hierarchy determination at $3\sigma$~C.L. for several detector configurations.}
\label{fig:mass}
\end{center}
\end{figure}

For CP-violation discovery, the best setup depends on the value of $\sin^22\theta_{13}$. The situation for mass hierarchy determination is completely different: the longer baseline represents the best solution. At the same OA 2.5$^o$, the Korean baseline gives an improvement in the sensitivity of a factor 5 with respect to the Kamioka baseline
(See Figure~\ref{fig:mass}). The higher rate, i.e. the smaller OA angle, improves the mass hierarchy determination:
going from OA~2.5$^o$ to OA~1$^o$ improves the sensitivity by a factor of 2.

\section{Other options at J-PARC and comparison to the CERN case}
\label{sec:cngs}

The main focus of the CERN program is the Large Hadron Collider (LHC)
however CERN is engaged in long baseline neutrino physics with
the CNGS project~\cite{CNGS} and supports T2K as a ``recognized'' experiment.
The CNGS beam has recently
begun operation and first events have been collected in the OPERA detector~\cite{Acquafredda:2006ki}.
The current optimization provides limited
sensitivity to the $\nu_\mu\rightarrow\nu_e$ reaction and OPERA should ultimately reach
a sensitivity $\sin^22\theta_{13} \lesssim 0.06$ (90\%C.L.) in 5 years of
running with the nominal $4.5\times 10^{19}$~pot/yr \cite{Komatsu:2002sz}.
The ICARUS~T600~\cite{t600paper}, still to be
commissioned, will detect too few contained CNGS events to competitively 
study electron appearance.

Ideas to improve the $\nu_\mu\rightarrow\nu_e$ sensitivity at
the CNGS have been discussed in the past~\cite{Ball:2006uw,Rubbia:2002rb}.
Recently, in Ref.~\cite{Meregaglia:2006du} we discussed the physics potential
of an intensity upgraded and 
energy re-optimized CNGS neutrino beam coupled to a 100~kton
liquid Argon TPC located at an appropriately chosen off-axis position,
and showed that improvements in $\theta_{13}$
sensitivity, search for CP-violation and mass hierarchy
determination were potentially possible. The discussion
relied on the observation that whereas J-PARC provides
a rapid cycle with high intensity proton bunches at $\sim 40$~GeV,
the CERN proton complex has fewer protons and a slower 
cycle but can accelerate up to 400~GeV. Hence, the 
resulting target beam powers are -- on paper -- comparable
(See Table~\ref{tab:accelerators}). In particular, it was noted
that future upgrades of the CERN LHC injection chain
(to be envisaged in the context of the luminosity upgrades) 
could provide increased proton intensities in
the SPS. This option labelled ``CNGS+" in Table~\ref{tab:accelerators}
accordingly envisioned $3.3 \times 10^{20}$ pots/yr.

The same idea was subsequently and independently re-analyzed assuming
a smaller detector of 20~kton 
located at an angle OA~0.8$^o$ at a baseline of 730~km 
 (MODULAr~\cite{Baibussinov:2007ea}).
In this case, two possible upgrades for CNGS beam labelled as ``CNGS1" and ``CNGS2"
yielding $1.2 \times 10^{20}$ pot/yr and $4.33 \times 10^{20}$ pot/yr
were considered.

A CERN accelerator division report~\cite{Meddahi:2007ju} subsequently indicated that
with an upgrade of the SPS RF and new injectors, it would indeed potentially be possible
to accelerate $2.4 \times 10^{20}$ pot/yr. This means that the CNGS+ 
exposure would correspond to a run of 7 years instead of 5 years 
and the CNGS2 beam to 9 years instead of the assumed 5 years.
Yet intensity limitations will be coming from the design of the equipment
in the current CNGS facility and from radiation and waste issues.
The desired intensities would therefore require a major re-assessment or a complete
reconstruction
of the CERN neutrino beam infrastructure (we however point out that the low
energy beams considered here would accommodate a significantly reduced decay tunnel
compared to the one of the CNGS).

In order to compare these options with the possible upgrades at J-PARC, we
focus on the $\sin^22\theta_{13}$ sensitivity. The obtained conclusions can
be readily extrapolated to the CP-violation and mass hierarchy determination.

\begin{figure} [tbh]
\begin{center}
     \includegraphics[width=.8\textwidth]{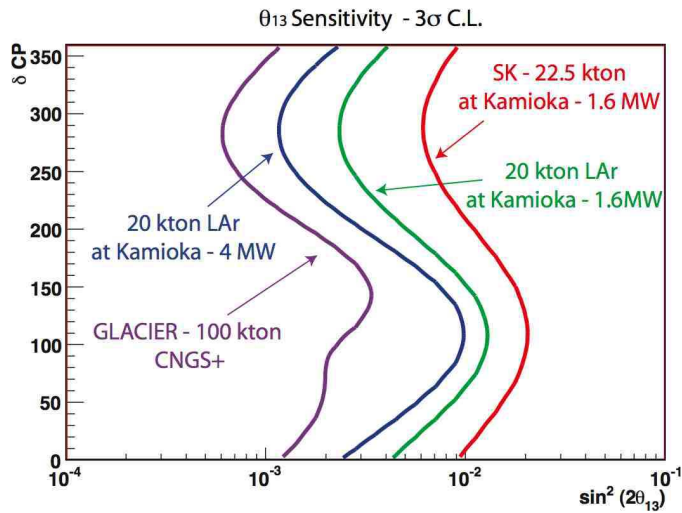}
\caption{$\theta_{13}$ sensitivity at $3\sigma$~C.L. for  20~kton LAr detector at 295~km, 2.5 degrees off-axis for 5 years of neutrino beam at 1.6~MW (green line) and 4~MW(blue line). For comparison the sensitivity of T2K (22.5~kton WC at Kamioka - 1.6~MW) and GLACIER-100~kton on upgraded CNGS~\cite{Meregaglia:2006du} are given.}
\label{fig:ALL1}
\end{center}
\end{figure}

\begin{figure} [tbh]
\begin{center}
     \includegraphics[width=.8\textwidth]{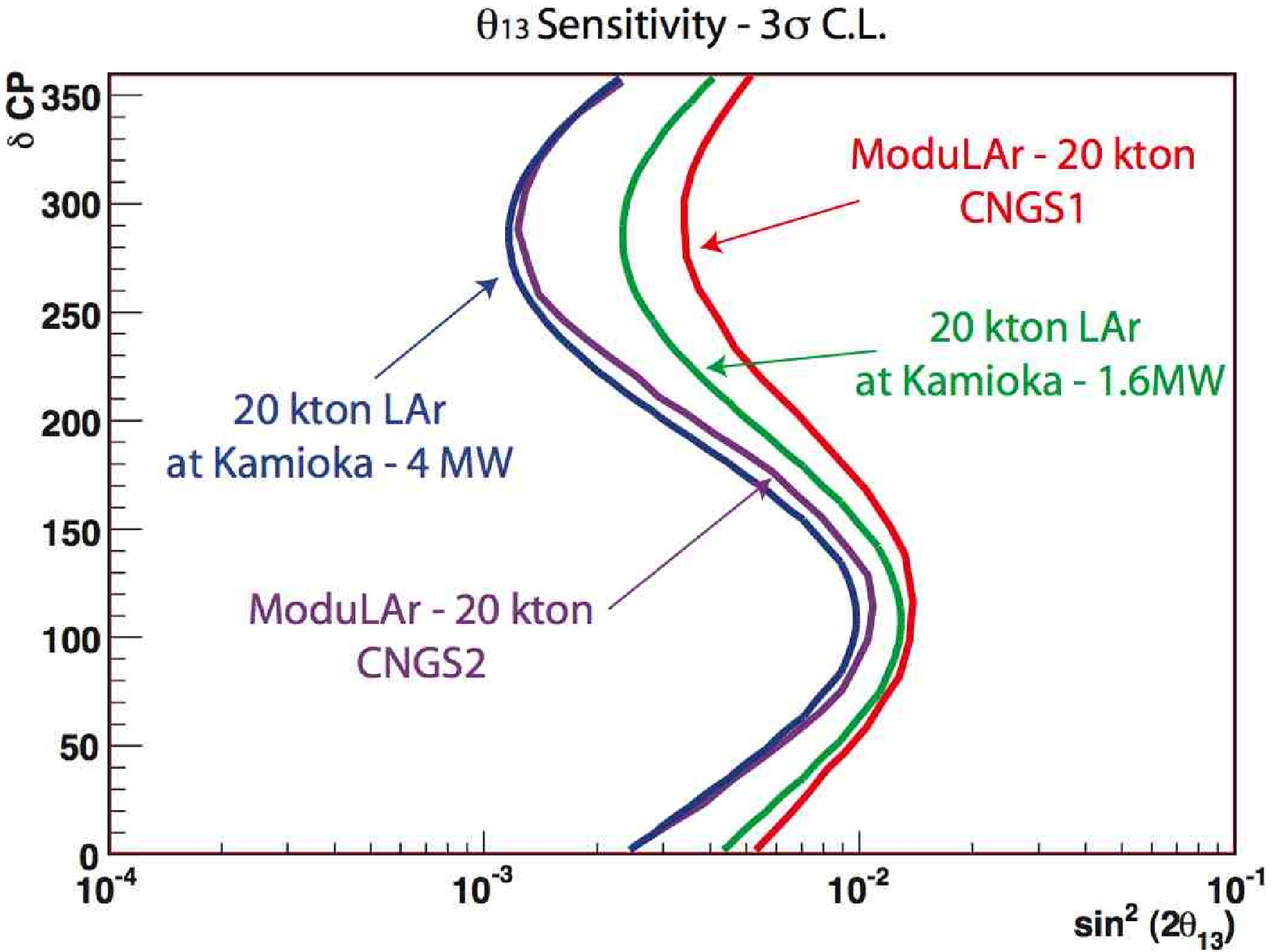}
\caption{$\theta_{13}$ sensitivity at $3\sigma$~C.L. for  20~kton LAr detector at 295~km, 2.5 degrees off-axis for 5 years of neutrino beam at 1.6~MW (green line) and 4~MW(blue line). For comparison the sensitivity of  ModuLAr-20~kton experiment~\cite{Baibussinov:2007ea} for two different upgrades of CNGS beam are shown (see text for details).}
\label{fig:ALL2}
\end{center}
\end{figure}

Figure~\ref{fig:ALL1} shows the expected $\theta_{13}$ sensitivity at
the $3\sigma$ C.L. for a 20~kton LAr TPC detector located at Kamioka
after 5 years of neutrino run with (a) the upgraded beam power of 1.6~MW
(b) the ultimate beam power of 4~MW. The expected sensitivity with the
existing SK detector and 1.6~MW is also shown. Finally, the sensitivity
of a 100~kton detector at the CNGS+ computed 
assuming 5 yeas of neutrino run and 5 years of antineutrino run 
is plotted. A 20~kton LAr TPC at
Kamioka is an effective way to improve the $\theta_{13}$-sensitivity
of the T2K experiment. 
Fig.~\ref{fig:ALL2} compares the sensitivity of a 20~kton LAr TPC at Kamioka to
the ModuLAr expectation~\cite{Baibussinov:2007ea}. 

As argued above, assuming similar
target masses,  the upgraded T2K and upgraded CERN beams theoretically provide
comparable sensitivities, however, we stress that the current CNGS beam line
infrastructure has not been designed to exceed $7.8\times10^{19}$ pot/yr, so further
upgrades and/or reconstructions must be envisaged in order to cope with the
potential increase of intensity from the SPS accelerator complex.

\section{An ``ultimate'' two-detectors configuration at J-PARC and the synergies with proton decay searches}
So far we have considered one far detector, either in the Kamioka region or in Korea.
In this last section, we discuss as 
``ultimate'' configuration the one which uses of two very large detectors, possibly
of different technologies, located at different baselines. 

\begin{figure} [p]
\begin{center}
     \includegraphics[width=.8\textwidth]{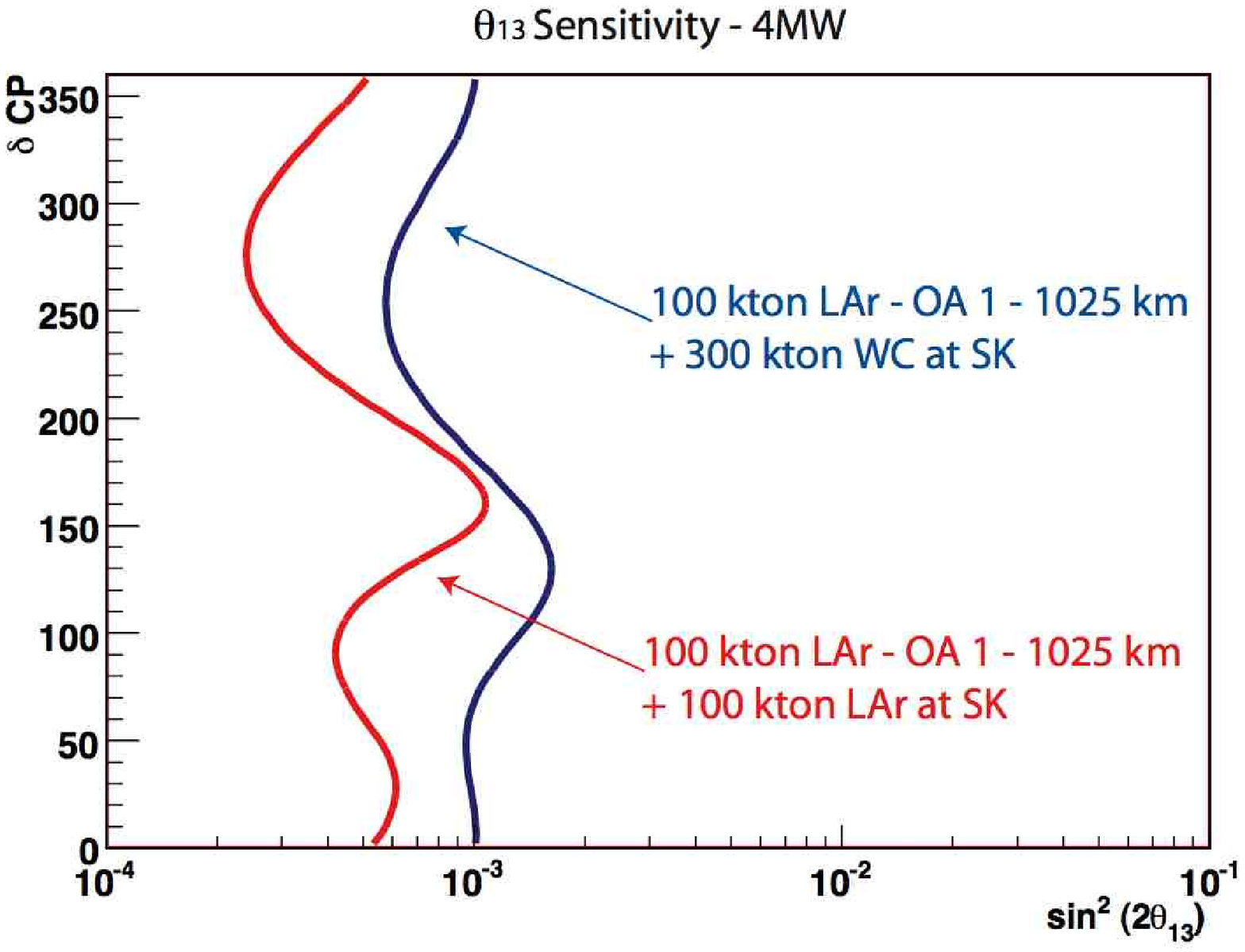}
\caption{$\theta_{13}$ sensitivity at $3\sigma$~C.L. for  a two detector configuration. A 100~kton LAr detector is located at 1025~km, 1 degree off-axis and a second detector i.e 300~kton WC (blue line) or 100~kton LAr (red line) is located at SK.}
\label{fig:th13FINAL}
\end{center}
\end{figure}

\begin{figure} [p]
\begin{center}
     \includegraphics[width=.8\textwidth]{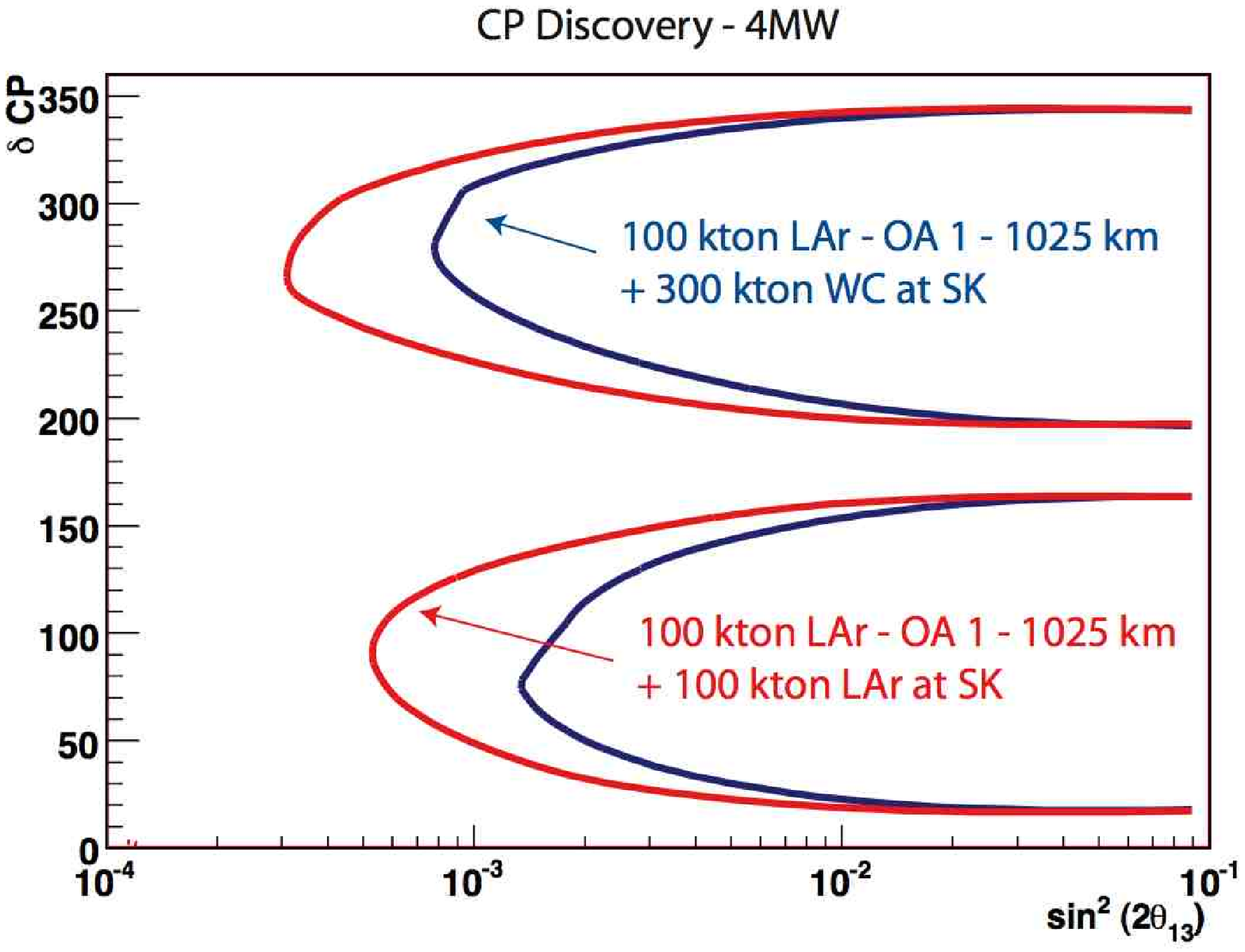}
\caption{CP-violation sensitivity at $3\sigma$~C.L. for  a two detector configuration. A 100~kton LAr detector is located at 1025~km, 1 degree off-axis and a second detector i.e 300~kton WC (blue line) or 100~kton LAr (red line) is located at SK.}
\label{fig:CPFINAL}
\end{center}
\end{figure}

In order to cope with the harder beam at the angle OA~1.0$^o$ we consider
a 100~kton LAr TPC at 1025~km in Korea. We have argued in the previous
paragraphs that the smaller off-axis angles give best performance for
the detector in Korea.
For the closer detector at 295~km OA~2.5$^o$ we consider
either a 300~kton Water Cerenkov detector or a 100~kton LAr TPC.
For the WC detector we used the ``standard'' analysis present in GLoBES as explained in appendix A of Ref.~\cite{Huber:2002mx} which treats QE and non-QE events in a different way analysing only the region between 0.4 and 1.2 GeV. 
The results on the sensitivity for $\theta_{13}$ and CP-violation are shown respectively in Fig.~\ref{fig:th13FINAL} and Fig.~\ref{fig:CPFINAL}. 

On one hand, following the argument of the authors of Ref.~\cite{Ishitsuka:2005qi}, the use of the same technology
for the detector at Kamioka and Korea would allow to cancel some instrumental systematic errors.
The option with two 100~kton LAr detector yields also better overall results, allowing for CP-violation 
discovery already for $\sin^2 (2 \theta_{13}) \sim 5 \times 10^{-4}$. 

On the other hand, if one considers the broader physics programme including also non-accelerator
physics as astrophysical neutrino observation (supernovae type II, etc...) and the search for proton
decay, studies show that the combination of water and Argon target would offer very attractive
complementarities in their physics programs (see e.g. Ref.\cite{Autiero:2007zj,Rubbia:2004yq}). As a concrete example, the
combination of the two technologies would allow to simultaneously address the $p\rightarrow e^+\pi^0$
and $p\rightarrow \bar\nu K^+$ channels with lifetime sensitivities above $10^{34}$~years.

\section{Conclusions}

Neutrino oscillations with one next generation liquid Argon TPC detector in Kamioka or in Korea
at an upgraded J-PARC neutrino beam offers very interesting prospects.
We concentrated on the physics reach of a 100~kton liquid Argon TPC. 
Several configurations were considered changing the baseline and the off-axis angle. 
If the detector is located at Kamioka (295~km and OA~2.5$^o$), 
a $3\sigma$ sensitivity for  $\sin^2(2 \theta_{13}) < 8 \times 10^{-4}$ could be achieved with
an 4~MW upgraded neutrino beam and 5 years of running.
Discovery of CP-violation at $3\sigma$ becomes possible down to $\sin^2(2 \theta_{13}) \sim 2 \times 10^{-3}$ 
for 50$\%$ of $\delta$ coverage.
If a signal will be observed in T2K, locating a detector in Korea is the best option to observe 
CP-violation, and the only option to discriminate between normal and inverted mass hierarchy.
A two-detector configuration with one at Kamioka and the other in Korea, 
although very challenging, would offer an even improved sensitivity
and very interesting complementarities, for example if two different detector technologies
were chosen.

\section*{Acknowledgments}
We are thankful to the organizers of the workshop, in particular Takaaki Kajita
and Soo-Bong Kim, for useful discussions
and the invitation to speak.
We acknowledge important exchanges with several people, in particular 
Takuya Hasegawa and Alberto Marchionni.
This work was supported by ETH/Zurich and the Swiss 
National Research Foundation. 



\end{document}